\documentclass[showpacs,aps,pre,numerical,twocolumn,10pt,superscriptaddress]{revtex4-1}
\usepackage{graphicx}
\usepackage{amsmath,amssymb}
\usepackage{times,mathptmx}
\usepackage{units}
\usepackage{epsfig}
\usepackage{booktabs}
\usepackage{multirow}

\usepackage[latin1]{inputenc}
\usepackage[english]{babel}
\usepackage{color}

\definecolor{darkgreen}{rgb}{0,0.5,0}
  
\begin{document}

\date{\today}
\title{Small Chimera States without Multistability in a Globally Delay-Coupled Network of Four Lasers}

\author{Andr\'e Röhm}\affiliation{Institut f{\"u}r Theoretische Physik, Technische Universit{\"a}t Berlin, 10623 Berlin, Germany}
\author{Fabian Böhm}\affiliation{Institut f{\"u}r Theoretische Physik, Technische Universit{\"a}t Berlin, 10623 Berlin, Germany}
\author{Kathy L{\"u}dge}\affiliation{Institut f{\"u}r Theoretische Physik, Technische Universit{\"a}t Berlin, 10623 Berlin, Germany}

\begin{abstract}
	We present results obtained for a network of four delay-coupled lasers modelled by Lang-Kobayashi-type equations. We find small chimera states consisting of a pair of synchronized lasers and two unsynchronized lasers. One class of these small chimera states can be understood as intermediate steps on the route from synchronization to desynchronization and we present the entire chain of bifurcations giving birth to them. This class of small chimeras can exhibit limit-cycle or quasiperiodic dynamics. A second type of small chimera states exists apparently disconnected from any region of synchronization, arising from pair synchronization inside the chaotic desynchronized regime. In contrast to previously reported chimera states in globally coupled networks, we find that the small chimera state is the only stable solution of the system for certain parameter regions, i.e. we do not need to specially prepare initial conditions.    	
\end{abstract}


\maketitle

\section{INTRODUCTION}
\label{secintro}

In the study of regular networks of identical phase oscillators it was believed until about ten years ago that the collective dynamics of all units would either lead to full synchronization or desynchronization\cite{PAN15}. However, Kuramoto and Battogtokh\cite{KUR02a} showed in 2002 that one could also obtain a mixed state containing both synchronized and desynchronized regions. This kind of state was later called 'Chimera State'\cite{ABR04}, after the hybrid creature of Greek legend. Significant efforts have since been made to understand these chimera states for networks of pure phase oscillators, and both analytical results in the continuum limit\cite{WOL11a, OME13a} as well as finite-size effects\cite{WOL11, PAN15b} have been obtained.  

There still exists no widely acknowledged formal definition for chimera states, even though some have been proposed\cite{ASH14, KEM16}. Nevertheless, a few unifying features are apparent that apply to most chimera states: Chimera states always include the simultaneous existence of a coherent or ordered subset of oscillators and an incoherent or unordered subset. Furthermore, the chimera state should exist even in the case when all oscillators are identical and the network possesses rotational symmetry, i.e.~the dividing of the system into ordered and unordered phase should be a dynamic effect, not one caused by inherent differences in the oscillators or spatial inhomogeneities.


The study of these seemingly arcane network patterns is of fundamental interest. The systems being studied are general and simple, so that knowledge of their patterns and the methods developed to understand them can be transferred to more complex real-world examples, e.g. neural networks in the brain, power grids or pacemaker cells in the human heart. Different oscillator models have been used as the fundamental units for the investigation of chimera states, among others the Kuramoto-type phase oscillators\cite{PAN15, KUR02a, ABR04, WOL11a, OME13a, WOL11, PAN15b}, coupled iterated maps \cite{OME11}, Stuart-Landau oscillators\cite{ZAK15b, SET13}, different excitable systems and neuronal models\cite{OLM11, OME13, HIZ13, VUE14a, OME15, ISE15b} and rotators\cite{OLM15}. Compounding the evidence for the generality of this kind of pattern, chimera states have been observed experimentally with chemical\cite{TIN12, NKO13}, electro-optical\cite{HAG12} or  mechanical oscillators\cite{MAR13} and in the electro-oxidation of silicon\cite{SCH14a}.

One of the systems that has been overlooked until recently are lasers. Lasers are a naturally suited system for the study of chimera and partial synchronization states. For one, lasers are networked all the time, e.g.~in the telecommunication business for optical data communication. Hence their stability when coupled, injected or exposed to delayed feedback is of vital interest. Furthermore, in the hierarchy of models, lasers represent a logical 'next step' after the pure Kuramato-type phase-oscillators, and oscillator models containing both variable phase and amplitude, e.g. Stuart-Landau, Fitzhugh-Nagumo, Van-der-Pol and Hopf-normal-form systems. Coupled laser equations contain both amplitude and phase, but in addition also the inversion of the gain medium, bringing the total dimensionality to 3. This also introduces an additional time-scale, as the gain inversion relaxes significantly slower compared to photons. Additionally, the propagation time between lasers can usually not be neglected in comparison to the internal dynamical time scales, so that delay-differential equations have to be used to describe the system.

While there has been an experimental demonstration of chimera states in a laser system\cite{LAR13}, these chimera states exist with respect to a virtual space variable and were found for a system consisting only of a single node with long delayed feedback. Generally, coupled lasers have been studied extensively in the past in the context of nonlinear laser dynamics\cite{KOZ00, VLA99, UCH01, YAN06b, DAH12, ARG16}, see also Ref.~\cite{SOR13} and references therein. But only recently has the connection with chimera states been made and a small chimera was numerically demonstrated in a network of four globally coupled lasers\cite{BOE15}. These small chimera states are qualitatively different from the Kuramoto-type phase-oscillator chimeras with respect to many previously established paradigms: They exist for very small network sizes, do not need a nonlocal coupling and are relatively robust with respect to the initial conditions. In this work we extend our previous work to larger feedback and coupling delays, which gives rise to new dynamics. We find small chimera states completely without the underlying multistability, i.e. as the only stable solution, contrasting other small chimera states found so far \cite{HAR16, BOE16, BOE15}. Because of this lack of multistability, we can directly identify the bifurcation sequence leading to their birth and death with varying the feedback and coupling strength. Furthermore, we even find a small chimera state with limit cycle dynamics. 


\section{MODEL}
\label{modelsec}


Our system consists of $Z = 4$ identical class~{B} lasers with self-feedback, modelled by Lang-Kobayashi-type differential equations\cite{LAN80b, ALS96}. The dynamical variables are the complex electric field amplitude $E$ and inversion $N$. We couple the lasers via their electric fields in an all-to-all coupling scheme with a delay time $\tau$, which in our case is always identical to the self-feedback delay time. This could, for example, be realized by symmetrically coupling all lasers into the same external cavity. The set of equations for the $n$th laser is given by: 

\begin{align}        
    \frac{dE_n}{dt}=&(1+i\alpha) E_n N_n + e^{-i C_p} \kappa 
    \sum_{j=1}^{Z} e^{ - i C_p} E_j ( t - \tau )
    \label{E-equation}\\
    \frac{dN_n}{dt}=&\frac{1}{T}(p-N_n-(1+2N_n)|E_n|^2) .
    \label{N-equation}
\end{align}

Here, the time has been renormalized with respect to the photon lifetime \cite{ALS96}. $T$ is the ratio of electron to photon lifetime, $\alpha$ is the amplitude-phase coupling and $p$ is the pump current. $\kappa$ is the strength of coupling and feedback, $C_p$ is the phase shift from the laser output to the external mirror. Note, that the same phase-shift also appears on the way from the mirror back to the laser. For this work, we assume identical phase shifts and delay-times for all lasers. The full feedback term is given by the sum of the delayed electric field signals of all lasers.  In this work we use $\alpha = 2.5$, $\tau = 40$, $p = 0.23$, $T = 392$ and $Z = 4$.

We numerically integrate Eq.~$\left(\ref{E-equation}\right)$ and $\left(\ref{N-equation}\right)$ with an Euler-type delay-differential equation solver written in C++. First, we simulate the time evolution for several thousand delay times to compensate for transient dynamics. Then we evaluate the dynamics and synchronization type of the resulting system state over one delay period, looking for differences between lasers. For this, we only compare the intensities $|E_n|^2$ of the lasers. Therefore we do not distinguish between in-phase ('synchronous') and anti-phase ('anti-synchronous') synchronization solutions\cite{YAN04c}. Almost all regions of synchronization discussed later in this work are synchronous, with the exception of a small region around $C_p \simeq \pi/2$. The bifurcation parameters are coupling phase $C_p$ and coupling strength $\kappa$. 

\section{Small Chimera States}

\begin{figure}[t]
  \includegraphics[width=0.45\textwidth]{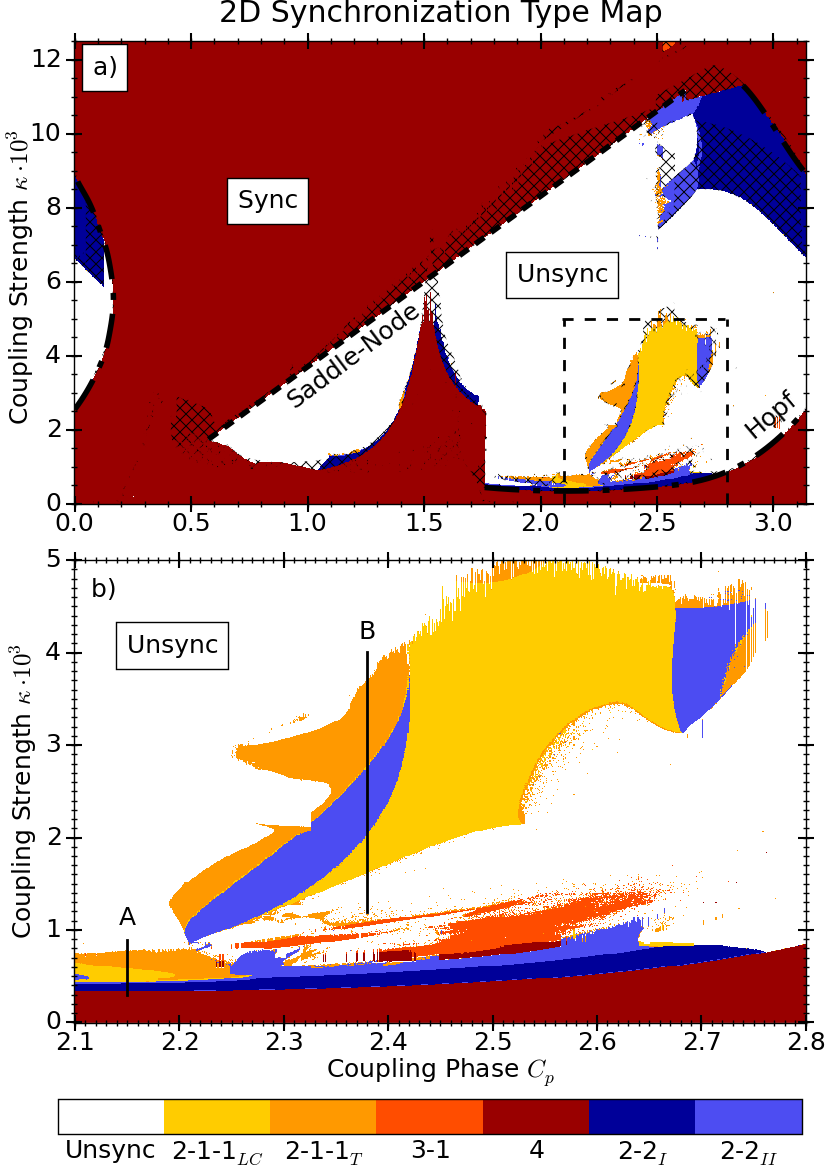}
  \caption{2D plot of the intensity synchronization type of our globally delay-coupled 4-laser network versus coupling strength $\kappa$ and coupling phase $C_p$. Panel b) is a zoom of the framed area of panel a). Synchronization state labelled '4' corresponds to full synchronization, 'Unsync' is the unsynchronized regime, other names describe the size of the partial clusters, e.g. '2-1-1' is a small chimera state. Regions of multistability are hatched in a). Parameters: $\alpha = 2.5$, $T = 392$, $\tau = 40$ and $p = 0.23$. }
  \label{2d_plot}
\end{figure}

In this work, we are looking for regions of partial synchronization and their generation mechanisms for the 4-laser network with intermediate delay, with a focus on small chimera states. We refer to the delay as 'intermediate', as the delay time $\tau = 40$ is large when compared to the photon lifetime $T_{photon} = 1$, but small compared to the electronic lifetime $T_{electron} = 392$ and relaxation oscillation period $T_{RO} \simeq 200$. We classify the type of synchronization pattern of the network with a series of integers corresponding to the size of clusters of synchronized lasers, going from the largest cluster to the smallest. Hence, full synchronization is labelled '4' in the case of $Z = 4$. Partial synchronization states include the triplet state '3-1', a cluster of 3 synchronized lasers and a solitary laser, the double-pair state '2-2' for two separate pairs of synchronized lasers and the small chimera state '2-1-1'. For the small chimera state the coherent region consists of the pair of synchronized lasers, while the incoherent domain is formed by the two remaining, desynchronized lasers. The unsynchronized regime is labelled 'Unsync'.

Figure~\ref{2d_plot} shows a two-dimensional plot for different coupling phases $C_p$ and coupling strengths $\kappa$ with colour coded synchronization types. The further subdivision of the partial synchronization states, e.g. '2-2$_I$' and '2-2$_{II}$', will be explained in more detail later. Fig.\,\ref{2d_plot}\,a) shows the results for coupling phases $C_p$ from 0 to $\pi$. Note, that due to the definition of the coupling term in Eq.~$\left(\ref{E-equation}\right)$ results are $\pi$-periodic in the coupling phase. The results were obtained numerically by linescans from high to low coupling strengths, i.e.~using the end state of the system of one linescan step as the initial condition for the next (sweep). To avoid lasers synchronizing on unstable states it proved necessary to apply small kicks of $\Delta |E_n| = 10^{-4}$ to the intensity in each step, with each laser being kicked differently to break apart unstable clusters. By varying the direction of the scan and using fixed initial conditions in additional numerical simulations we could identify some regions of multistability, which are hatched in Fig.\,\ref{2d_plot}\,a). 

The shape of the synchronized region (brown) in Fig.\,\ref{2d_plot}\,a) is determined by two effects. First, as explained in \cite{YAN04c, BOE15}, the parameter space of $C_p$ and $\kappa$ is organized by an Andronov-Hopf and a pitchfork bifurcation originating from the effects of coupling two or more lasers without time-delay. In our case the boundaries are similar in nature, yet deformed by the significant delay-times. One Andronov-Hopf-Bifurcation is indicated by the dashed-and-dotted line on the lower right of the desynchronized (white) area in Fig.\,\ref{2d_plot}, while there is also another one on the left side of the synchronization tongue (brown) at $C_p \simeq \pi/2$. Second, we also see some similarity to the bifurcations of a single laser with delay. The straight, upper left boundary of the desynchronized (white) area marked by the black dashed line in Fig.\,\ref{2d_plot}\,a) is caused by the delay-dynamics. It corresponds to the birth of the next external cavity mode (ECM) in a saddle-node bifurcation\cite{LYT97, PIE01} of a single laser with feedback. For our parameters the lasers are always operating in the continuous wave mode inside the continuous brown synchronized regime, i.e.~their output intensities are constant and identical. Contrastingly, the lasers exhibit chaotic or quasiperiodic behavior inside the large continuous white region of desynchronization, with a few exceptions of higher order limit cycles. On the boundaries between synchronization and desynchronization and inside the desynchronized regime we find a multitude of partial synchronization states, that will be discussed in more detail.

Fig.\,\ref{2d_plot} b) shows a zoom of the framed area in Fig.\,\ref{2d_plot}\,a) with many partial synchronization states (coloured regions). This region is of special interest, as it contains the small chimera states '2-1-1' (orange/yellow). We find transitions between almost any two types of partial synchronization, indicating a large zoo of bifurcations. Notably, the chimera state '2-1-1' and double-pair state '2-2' are significantly more prevalent for these parameters than the triplet state '3-1'. Only a small band of synchronization (brown) stretches across the lower edge of Fig.\,\ref{2d_plot}\,b). The transition from synchronization to desynchronization greatly varies with the coupling phase $C_p$, however some general trends are visible. First, the only partial synchronization state directly neighbouring the cw-synchronization band (brown) on the bottom is the symmetric double-pair state '2-2$_{I}$' (dark blue), a 2-cluster state with time-periodic intensity dynamics. The transition is given by the boundary, where a corresponding system of two coupled lasers desynchronizes in an Andronov-Hopf bifurcation\cite{YAN04c}, giving rise to periodic intensity fluctations. Second, and contrastingly, we find that desynchronization can border any other solution, i.e. there exist abrupt transitions to desynchronization. Third, there exists multistability around $C_p \simeq 2.4$ and $\kappa \simeq 0.7 \cdot 10^{-3}$, as can be seen by the broken up nature of synchronized areas and 'stripes', caused by the direction of the underlying linescans.  

We are interested in the small chimera states and their connections to the other surrounding partial synchronization states. Therefore we will present two different types of creation mechanisms of the small chimera states '2-1-1'. The black lines labelled \textbf{A} and \textbf{B} in Fig.\,\ref{2d_plot}\,b) will be explored as linescans and the different system states will be discussed in more details in the following sections. 

\section{Transition to desynchronization}

\begin{figure}[tbh]
  \includegraphics[width=0.45\textwidth]{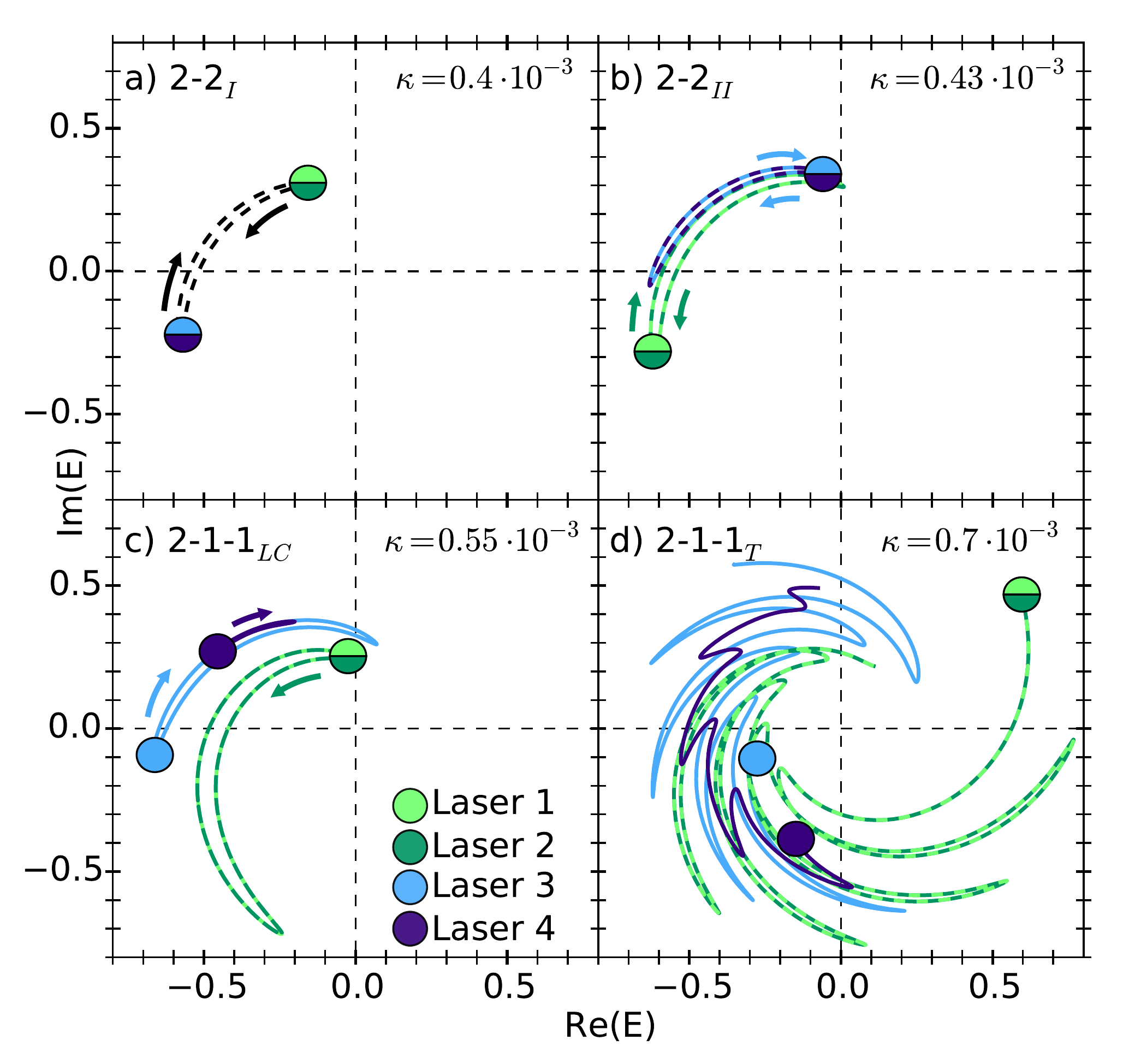}
  \caption{Projection of dynamic states into the plane of the complex electric field amplitude $E$ for different coupling strengths $\kappa$. Coloured circles represent simultaneous positions of the different lasers in the network. Lines in a)-c) depict the outline of limit cycles when corrected for a constant phase velocity (rotating frame). Subplot d) shows part of the quasi-periodic small chimera state '2-1-1$_T$' trajectory. All states lie along linescan \textbf{A} as shown in Fig.\,\ref{linescan_chimeras} and are labelled identically. Coupling strengths are shown in the upper right corners. Timeseries for the same states are shown in Fig.\,\ref{linescan_chimera_timeseries}. Parameters: $C_p = 2.15$, $\alpha = 2.5$, $T = 392$, $\tau = 40$ and $p = 0.23$. }
  \label{linescan_chimera_phasesnapshots}
\end{figure}

For very small coupling and feedback strengths $\kappa$, the system is in the fully synchronized continuous wave state, i.e. all lasers are frequency and phase-locked with constant identical intensities. However, as can been seen in Fig.\,\ref{2d_plot}~a) the system desynchronizes for higher $\kappa$ for certain coupling phases $C_p$. This transition can be abrupt or contain a multitude of partial synchronization states. The transition marked by line~\textbf{A} in Fig.\,\ref{2d_plot}~b) crosses the largest number of partial synchronization states, including small chimera states, and will be explored in more detail in this section.

Before discussing the transitions, we want to first characterize different partial synchronization states that are present in the network of four lasers. For this, Fig.\,\ref{linescan_chimera_phasesnapshots} depicts exemplary snapshots and periodic limit cycles for stable states found along the line~\textbf{A} of Fig.\,\ref{2d_plot}~b) projected onto the plane of the complex electric field amplitude $E$. Here, the complex electric field amplitude $E$ has been plotted in the same plane for all four lasers. Where possible, we have corrected the rotating frame with a constant phase velocity to prevent limit cycles turning into quasiperiodic solutions. 

During evaluation of the time dependent dynamics we have discovered that it is useful to further subdivide the partial synchronization states. Fig.\,\ref{linescan_chimera_phasesnapshots}\,a) and b) depict the difference between double-pair states '2-2$_I$' and '2-2$_{II}$', with parameters of linescan~\textbf{A} in Fig.\,\ref{2d_plot}\,b). Both states can be characterized by two pairs of synchronized lasers, where the dynamics are out-of-phase. The symmetric double-pair state '2-2$_I$' case is characterized by both clusters moving along the same limit cycle, but shifted by half a period. Contrastingly, in the asymmetric double-pair state '2-2$_{II}$' the dynamics of both clusters are different as projected in Fig.\,\ref{linescan_chimera_phasesnapshots}\,b) (compare green and blue limit cycles). The two pairs do not exhibit the same limit cycle. We also subdivide the small chimera states into two classes: Fig.\,\ref{linescan_chimera_phasesnapshots}\,c) depicts the small chimera state with periodic intensity dynamics '2-1-1$_{LC}$'. Because only a single pair of lasers is synchronized, three different limit cycles are present (blue, green and violet). The small chimera state with quasiperiodic intensity dynamics '2-1-1$_{T}$', i.e.~phase space trajectories on a torus, is shown in Fig.\,\ref{linescan_chimera_phasesnapshots}~d). For the phase space projection shown in Fig.\,\ref{linescan_chimera_phasesnapshots}\,d) no attempt to correct the rotating frame was made and only a small section of the time series is shown. In all the cases shown here, the long-term average phase velocity for all four lasers is identical. This is, however, not generally true for all states in the system. 

\begin{figure}[tbh]
  \includegraphics[width=0.45\textwidth]{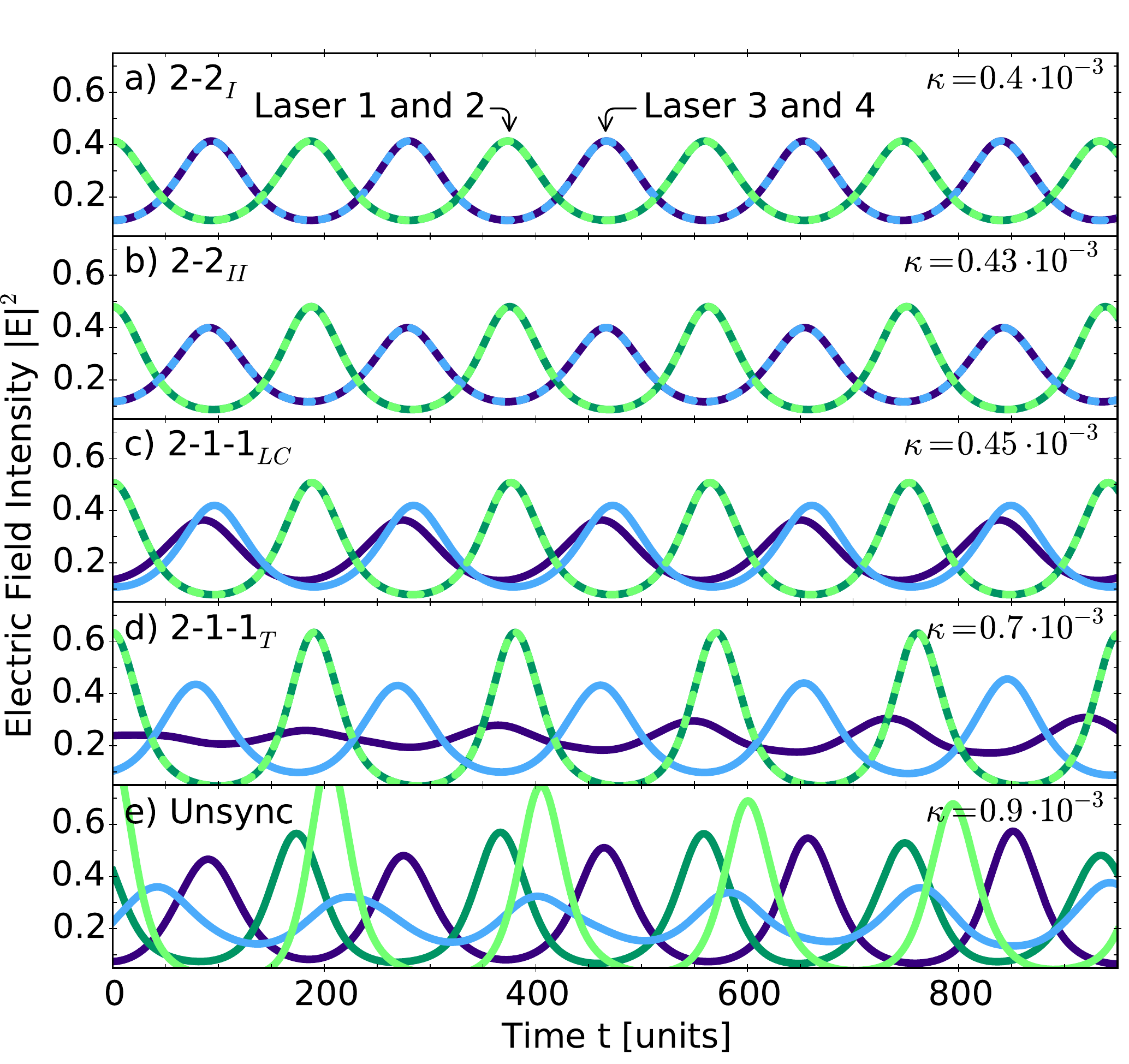}
  \caption{Aligned time series slices of the electric field intensity $|E|^2$ for the linescan A indicated in Fig.\,\ref{2d_plot}\,b) and shown in Fig.\,\ref{linescan_chimeras}, corresponding pictures of snapshots shown in Fig.\,\ref{linescan_chimera_phasesnapshots}. Coupling strengths are shown in the upper right corners. Parameters: $C_p = 2.15$, $\alpha = 2.5$, $T = 392$, $\tau = 40$ and $p = 0.23$. }
  \label{linescan_chimera_timeseries}
\end{figure}

To further illustrate the evolution of the network dynamics, Fig.\,\ref{linescan_chimera_timeseries} shows aligned sample slices of the time series for the different regions of linescan~\textbf{A}. In this visualization we can see the identical limit cycles of the symmetric double-pair state '2-2$_I$' in a), the different amplitudes of the asymmetric double-pair state '2-2$_{II}$' in b) and the periodic small chimera state dynamics '2-1-1$_{LC}$' in c). Once again, only a small section of the time series is shown for the quasiperiodic small chimera state '2-1-1$_T$' in Fig.\,\ref{linescan_chimera_timeseries}~d) and the unsynchronized state in Fig.\,\ref{linescan_chimera_timeseries}~e). 

\begin{figure}[tbh]
  \includegraphics[width=0.45\textwidth]{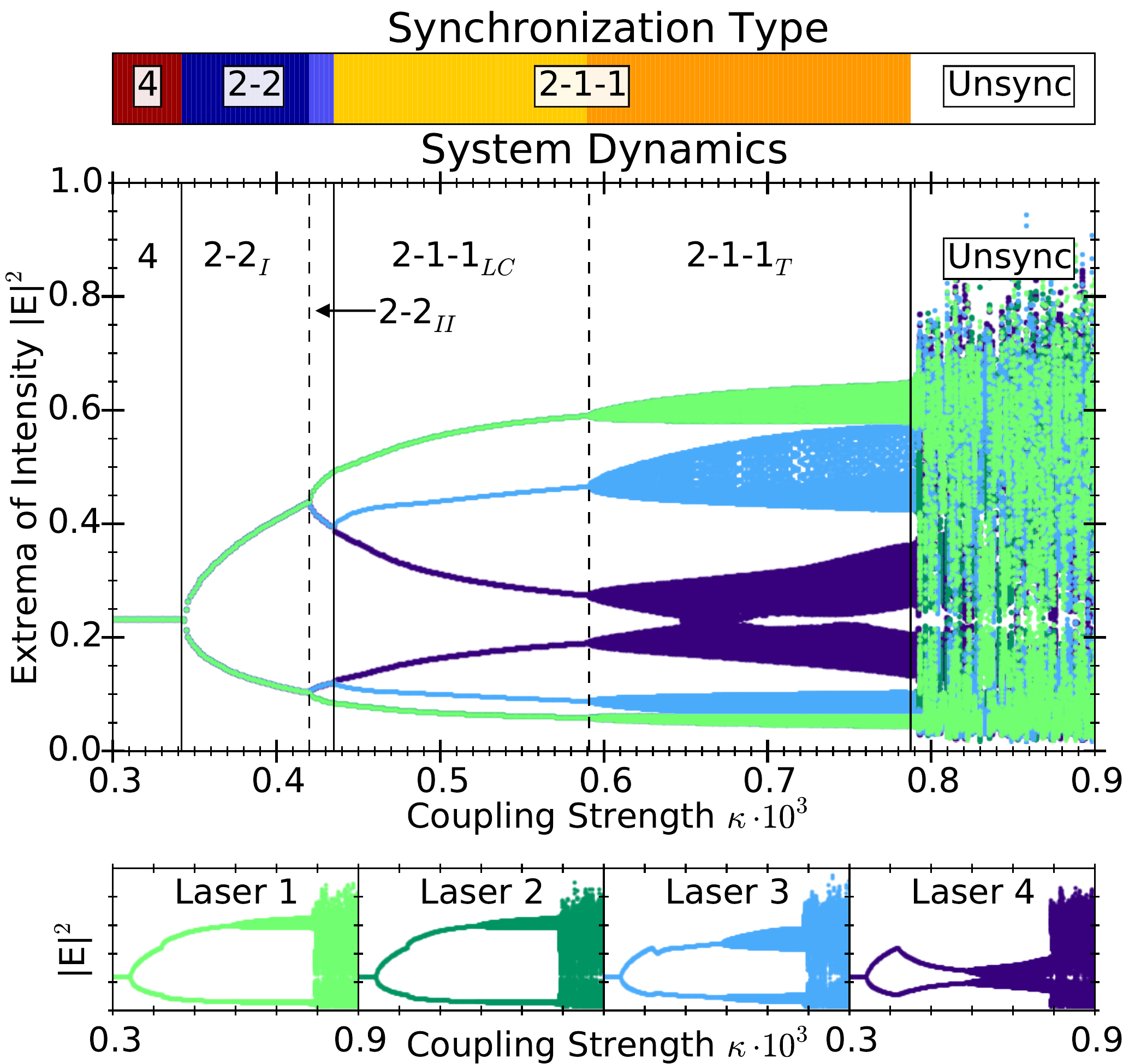}
  \caption{Bifurcation diagram of linescan \textbf{A}: Extrema of the electric field intensity $|E|^2$ of the 4 lasers plotted against coupling strength $\kappa$ along Linescan~\textbf{A} in Fig.\,\ref{2d_plot}\,b). The top panel depicts the synchronization type in colour code, the bottom insets show the same linescan data split for the individual lasers. Parameters: $C_p = 2.15$, $\alpha = 2.5$, $T = 392$, $\tau = 40$ and $p = 0.23$. }
  \label{linescan_chimeras}
\end{figure}

To better represent the transitions from one state to the next, we will now study the bifurcation diagram for the linescan \textbf{A}. Fig.\,\ref{linescan_chimeras} shows the extrema of the electric field intensity $|E|^2$ of all lasers versus coupling strength $\kappa$. The top panel and text labels denote the synchronization type of the lasers, with vertical lines at the bifurcation points. The row of small insets on the bottom shows the extrema for each laser individually, to reveal synchronized pairs. We find a multitude of bifurcations, altering both the system dynamics and synchronization patterns. For small coupling strengths the network emits continuous wave and is fully synchronized. Consequently the system only possesses a single extremum in Fig.\,\ref{linescan_chimeras}. At $\kappa \simeq 0.34 \cdot 10 ^{-3}$ the system undergoes a symmetry-breaking Andronov-Hopf bifurcation marked by the first black vertical line. The system then changes to the symmetric double-pair state '2-2$_{I}$'. Fig.\,\ref{linescan_chimera_timeseries}\,a) shows the intensity time series for this type of synchronization pattern. Like the phase-portrait, it is characterised by two pairs of synchronized lasers with identical, but out-of-phase limit cycles, cf.~Fig.\,\ref{linescan_chimera_phasesnapshots}~a). This kind of dynamics has been called 'anti-phase self-pulsing' by the authors of Ref.~\cite{KOZ00} and according to their results is typical for the desynchronization of a laser ensemble for small delay times $\tau$. The frequency of these intensity oscillations is almost identical to the relaxation oscillation (RO) frequency. This is consistent with previous results for feedback-induced RO-undamping\cite{ERN00a, LYT97, KOZ00}. Furthermore, the frequency of oscillations for all subsequent states does not significantly change for higher coupling strengths $\kappa$, as can be seen by the alignment of time series in Fig.\,\ref{linescan_chimera_timeseries} a)-d). One can assume that this is in part caused by the delay being short with respect to the carrier equation lifetime, as this prevents the system from sustaining oscillations with the period of the delay time $\tau$.

To the right of the first dashed vertical line in the bifurcation diagram Fig.\,\ref{linescan_chimeras} at $\kappa \simeq 0.42 \cdot 10 ^{-3}$ the extrema of the two clusters diverge. Yet, as can be seen in the bottom row each individual laser still possesses only two extrema, therefore the apparent splitting of the line of extrema in the centre panel of Fig.\,\ref{linescan_chimeras} does not correspond to a period-doubling. Rather, it is the transition from the symmetric double-pair state '2-2$_I$' to the asymmetric double-pair state '2-2$_{II}$' via symmetry breaking. Once again, cf. Fig.\,\ref{linescan_chimera_timeseries}\,b) and Fig.\,\ref{linescan_chimera_phasesnapshots}\,b) for the time evolution of the asymmetric double-pair state '2-2$_{II}$' solution: The period of the oscillations is still matched and the two clusters' oscillations are still time-shifted by half an oscillation period. However, the limit cycles are no longer identical. 

The small chimera states '2-1-1$_{LC}$' appear for $\kappa \geq 0.44 \cdot 10^{-3}$ and can be seen as a further subdivision of extrema in Fig.\,\ref{linescan_chimeras}. These chimera states evolve from the asymmetric double-pair state '2-2$_{II}$' when the smaller amplitude cluster breaks apart, without disturbing the larger amplitude cluster. Fig.\,\ref{linescan_chimera_timeseries}\,c) and Fig.\,\ref{linescan_chimera_phasesnapshots}\,c) show the time evolution after the formation of the small chimera state with periodic limit-cycle dynamics. First the limit cycles of the two newly desynchronized lasers are still quite similar to each other, but further diverge for increasing coupling strengths. As long as $\kappa \leq 0.59 \cdot 10^{-3}$ the chimera state is exhibiting periodic intensity oscillations. At $\kappa \simeq 0.59 \cdot 10^{-3}$ the entire system undergoes what we assume to be a secondary Andronov-Hopf bifurcation, without changing the synchronization state, i.e. without breaking the synchronization within the 2-laser cluster. The resulting small chimera state with quasiperiodic dynamics '2-1-1$_{T}$' exhibits the typical slow intensity modulation, as can be seen by its time series shown in Fig.\,\ref{linescan_chimera_timeseries}\,d). The corresponding phase-snapshot is shown in Fig.\,\ref{linescan_chimera_phasesnapshots}\,d). For coupling strengths $\kappa \geq 0.79 \cdot 10^{-3}$, corresponding to the rightmost black line in  Fig.\,\ref{linescan_chimeras}, the system leaves the chimera state and is fully desynchronized afterwards. The time series is shown in Fig.\,\ref{linescan_chimera_timeseries}\,e). 

The authors of~\cite{PEC14} in detail discuss the possibility of some clusters in a network desynchronizing without disturbing the synchronization of other clusters. The authors refer to this as 'isolated desynchronization' (ID) and the transition from the asymmetric double pair state '2-2$_{II}$' to the small chimera state '2-1-1$_{LC}$' in Fig.\,\ref{linescan_chimeras} for our system can be seen as such an isolated desynchronization. However, the authors of~\cite{PEC14} are mostly concerned with clusters given by irregular networks. In our globally coupled network the subdivision into two pairs of synchronized lasers is a dynamical effect not predetermined by the topology. Furthermore, we want to emphasize that in our system the larger-amplitude cluster of the asymmetric double-pair state '2-2$_{II}$' not only persists after the creation of the small chimera state, but also survives the transition from periodic to quasi-periodic dynamics.

On the whole, linescan~\textbf{A} represents the transition from synchronization to desynchronization with increasing coupling and self-feedback strength. Starting from the fully synchronized state, the system first undergoes a symmetry-breaking Andronov-Hopf-Bifurcation, typical for laser ensembles, and then loses the symmetry of the resulting laser pairs in another symmetry-breaking bifurcation. One of these synchronized pairs then breaks apart without disturbing the other via 'isolated desynchronization'. The system then undergoes a secondary Andronov-Hopf-Bifurcation, before reaching the fully desynchronized state. For all the states found in between we tested the system for multistability by changing the sweep-direction of the linescan and also performed a systematic scan of a reasonably large subspace of initial conditions for selected points. From this we strongly believe there to be no multistability in this linescan. 

The existence of small chimera states in networks of globally coupled oscillators was first reported in \cite{BOE15}. Recently small chimera states have also been found experimentally in a network of four globally-coupled chaotic opto-electronic oscillators\cite{HAR16}. Yet, in both cases they were assumed to be linked with a high degree of multistability. Contrastingly, we find no multistability for the linescans \textbf{A} and \textbf{B} (shown in the following section). It seems very intriguing, that the only stable synchronization type for our network is such a seemingly arbitrary, asymmetric state for large regions in parameter space. This also sets our results apart from previously obtained results for partial synchronization patterns in rings of chaotic Rössler systems\cite{HU00a}. Followingly, after a time of transient dynamics the system always reaches the small chimera state and we do not need to specially prepare initial conditions, as is necessary in the Kuramoto-type phase oscillator chimera states\cite{ABR04}. The small chimera state arises from the synchronous domain after three different bifurcations: First an Andronov-Hopf, then a symmmetry-breaking bifurcation and finally the isolated desynchronization of one of the laser pairs. Contrastingly, only a single bifurcation separates it from the fully desynchronized regime.


\section{Island of partial synchronization}

\begin{figure}[tbh]
  \includegraphics[width=0.45\textwidth]{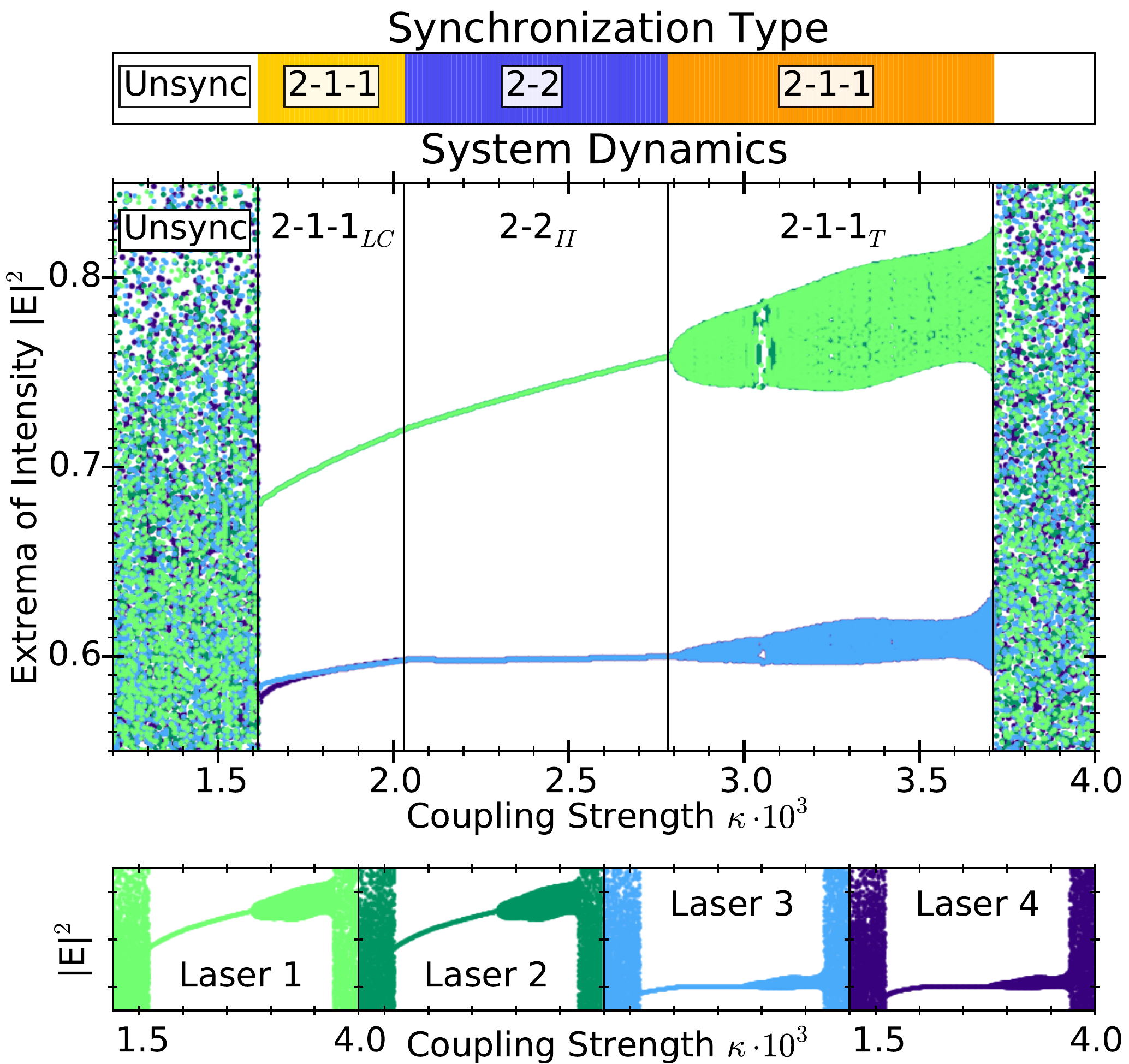}
  \caption{Bifurcation diagram of linescan~\textbf{B}: Extrema of the electric field intensity $|E|^2$ of the 4 lasers plotted against coupling strength $\kappa$. The coupling phase was kept constant $C_p = 2.38$, corresponding to Linescan \textbf{B} in Fig.\,\ref{2d_plot} b). The top panel depicts the synchronization type in colour code, the bottom insets show the same linescan data split for the individual lasers. Parameters: $\alpha = 2.5$, $T = 392$, $\tau = 40$ and $p = 0.23$. }
  \label{additional_linescan}
\end{figure}

Apart from the gradual desynchronization discussed in the previous section, a second region of small chimera states (orange/yellow) is visible in the upper part of Fig.\,\ref{2d_plot}\,b). They exist in an isolated region of partial synchronization states inside the desynchronized (white) regime. They therefore do not represent intermediate steps on the route from synchronization to desynchronization. These small chimera states and their bifurcations behave qualitatively different to the ones discussed in the previous section, spontaneously arising from the chaotic, fully desynchronized regime.

\begin{figure}[tbh]
  \includegraphics[width=0.45\textwidth]{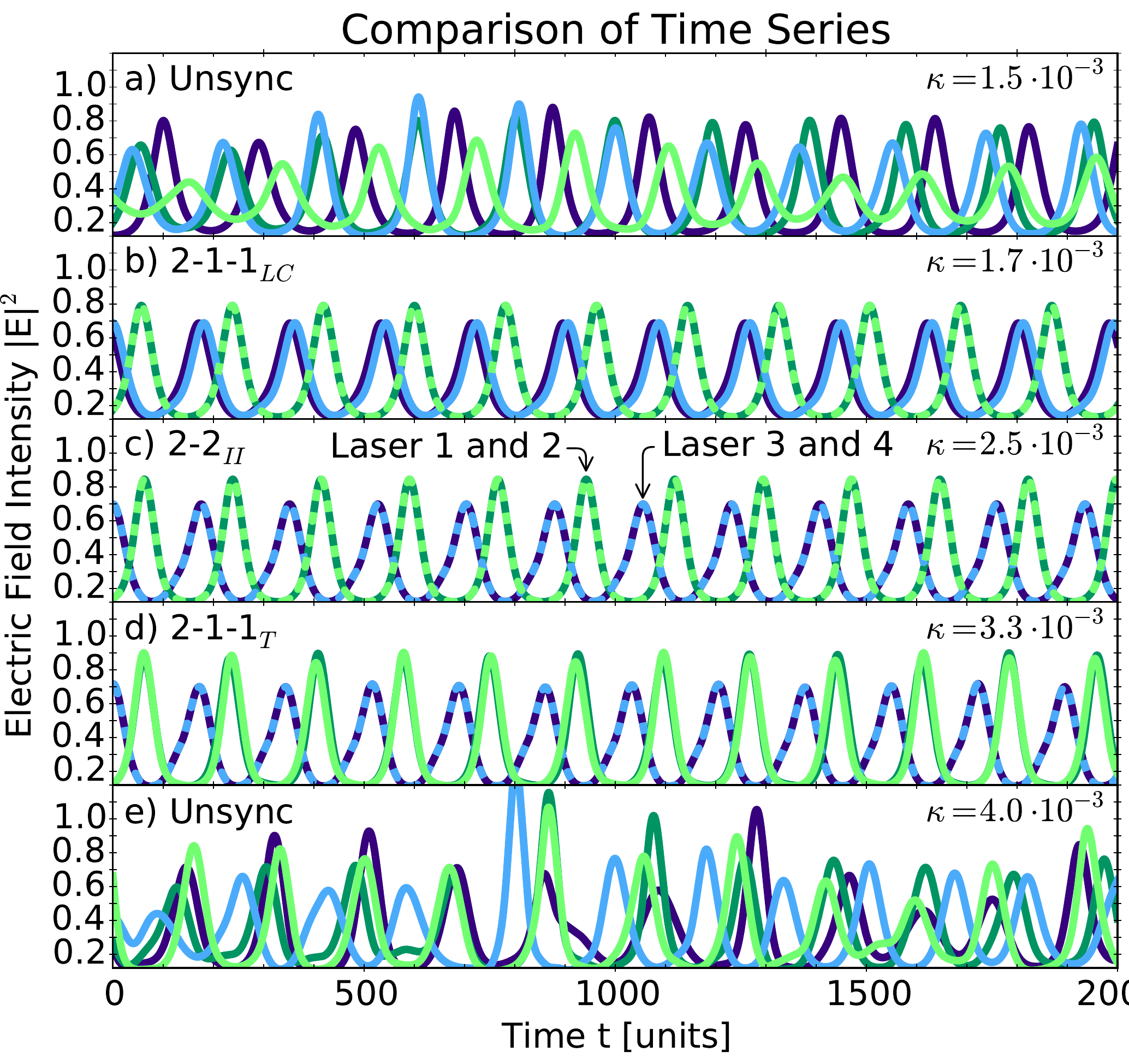}
  \caption{Aligned time series slices of the electric field intensity $|E|^2$ for the linescan \textbf{B} indicated in Fig.\,\ref{2d_plot}\,b) and shown in Fig.\,\ref{additional_linescan}. Coupling strengths are shown in the upper right corners. Parameters: $C_p = 2.38$, $\alpha = 2.5$, $T = 392$, $\tau = 40$ and $p = 0.23$. }
  \label{additional_linescan_timeseries}
\end{figure}

As done for the previous section, we present the bifurcation diagram for the second linescan, labelled \textbf{B} in Fig.\,\ref{2d_plot}\,b) with $C_p = 2.38$. Fig\,\ref{additional_linescan} depicts the extrema for the electric field intensities of the four lasers. Note the scaling of the y-axis chosen in Fig.\,\ref{additional_linescan} for greater detail. The corresponding time series are shown in Fig.\,\ref{additional_linescan_timeseries}. At coupling strength $\kappa \simeq 1.6 \cdot 10^{-3}$ the system changes from the unsynchronized regime to a small chimera state with limit cycle dynamics '2-1-1$_{LC}$'. Note, that laser 3 and 4 are the desynchronized lasers, with only a small difference in their extrema that is hard to spot at the scale shown in Fig.\,\ref{additional_linescan}, but can inferred from the time series in Fig.\,\ref{additional_linescan_timeseries}~b), which is not a transient. This transition is noteworthy, as it directly links the small chimera state with limit cycle dynamics '2-1-1$_{LC}$' with the quasiperiodic, unsynchronized regime without any intermediate steps. For increasing coupling strength $\kappa$ the limit cycles of laser 3 and 4 approach each other and synchronize, resulting in the asymmetric double-pair state '2-2$_{II}$' at $\kappa \simeq 2.0 \cdot 10^{-3}$. The time series shown in Fig.\,\ref{additional_linescan_timeseries}~c) looks similar to the one shown for linescan \textbf{A} in Fig.\,\ref{linescan_chimera_timeseries}\,b), however the asymmetry between clusters is even larger here. Going towards higher coupling strengths, this state loses stability at $\kappa \simeq 2.8 \cdot 10^{-3}$, marked by the third black vertical line in Fig.\,\ref{additional_linescan}. Here, lasers 1 and 2 lose synchronization, while lasers 3 and 4 stay synchronized, as opposed to the break-up of the cluster made from laser 3 and 4 seen below $\kappa \simeq 2.0 \cdot 10^{-3}$. In this context, note that while all nodes in our system are identical and could therefore theoretically be labelled in any order, the two clusters of synchronized laser pairs are clearly distinguishable by their different amplitudes. The asymmetric double-pair state '2-2$_{II}$' shown in the centre of Fig.\,\ref{additional_linescan} is therefore bounded by the breakup of one of the two different clusters on each side.

The small chimera state '2-1-1$_T$' appearing for higher coupling strengths $\kappa \geq 2.8 \cdot 10^ {-3}$ exhibits quasiperiodic behaviour. It is born in a bifurcation that changes the system dynamics and synchronization pattern at the same time. However, this chimera state is different when compared with the quasiperiodic chimera state '2-1-1$_{T}$' in Fig.\,\ref{linescan_chimeras}. The quasiperiodic chimera state '2-1-1$_{T}$' shown here for linescan \textbf{B} in Fig.\,\ref{additional_linescan} is not immediately identifiable as a chimera state in the bifurcation diagram. Lasers 1 and 2 do not show any divergence in the distribution of their extrema. Only the time series shown in Fig.\,\ref{additional_linescan_timeseries}~d) easily reveals the chimera state for $\kappa \geq 2.8 \cdot 10^{-3}$. This property can be explained by the fact that the small chimera state with quasiperiodic dynamics '2-1-1$_{T}$' for linescan \textbf{A} in Fig.\,\ref{linescan_chimeras} is born from a small chimera state with limit cycle dynamics '2-1-1$_{LC}$', i.e.~from a state where the desynchronized lasers are already well separated in amplitude. Consequently, it is born in a bifurcation that does not break up any clusters. On the other hand, the chimera state for linescan \textbf{B} is directly born from the asymmetric double-pair state '2-2$_{II}$' and therefore the extrema of laser~1 and laser~2 seem to overlap, as they arise from the same limit-cycle. This small chimera state loses its stability for even higher coupling strengths, transforming back into the desynchronized state.

Investigating the similarity between the trajectories of laser~1 and laser~2 in the quasiperiodic small chimera state '2-1-1$_{T}$' of linescan~\textbf{B} further by Poincar\'e sections, we found that the phase space volumes occupied by laser~1 and laser~2 seem to be identical. Hence, under the correct rotating frame, the desynchronized lasers seem to follow the outline of the same torus in phase space, but they are never at the same spot on the torus. Keeping in mind that it takes an infinite amount of time to actually fill out the entire torus surface in phase space, one could speculate that the two lasers do not exactly repeat each other in any finite amount of time. This is not testable numerically, however, due to finite numerical resolution limits. 

As a whole, linescan~\textbf{B} can be understood as the overlap of two regions of pair synchronization that give birth to the small chimera states. While Laser~1 and 2 are synchronized for $1.6 \cdot 10^{-3} \lesssim< \kappa \lesssim 2.8 \cdot 10^{-3}$, Lasers~3 and 4 synchronize for $2.0 \cdot 10^{-3} \lesssim \kappa \lesssim 3.7 \cdot 10^{-3}$. Where these two bands overlap, we recover the asymmetric double-pair state '2-2$_{II}$'. However, the type of synchronization is different in these two bands, as we find limit cycle chimeras '2-1-1$_{LC}$' on the one side, and quasiperiodic chimeras '2-1-1$_T$' on the other side. As mentioned, we did not find any multistability in this region.

%
%
%


\section{Conclusions}
\label{sec_Conclusion}

We investigated the dynamics and synchronization patterns of a 4-laser network with intermediate coupling delay. We found several regions of partial synchronization, with small chimera states embedded in different bifurcation scenarios. One identified scenario for the birth of a small chimera state lies on the route to desynchronization induced by symmetry-breaking Andronov-Hopf bifurcation and a subsequent isolated desynchronization transition. The second route found is pair synchronization happening within the region of chaotic unsynchronized dynamics. Somewhat surprisingly, as opposed to previously reported chimera states in small globally coupled networks, these states found here do not exhibit any underlying multistability, excluding the trivial symmetry of permutations of the lasers. We therefore find large regions in parameter space where the only solution for our network (with identical lasers, coupled completely symmetrically) are very asymmetric partial synchronization states, independent from the chosen initial conditions, especially also for initial conditions very close to the fully synchronized state.

Naturally, the question now arises how these findings will transfer to networks with more coupled lasers. Preliminary results indicate that the general shape of the partial synchronization regions remains similar, but a complete answer will be the topic of further studies. 


\section{Acknowledgement}

The authors thank B. Lingnau and L. Jaurigue for fruitful discussions. This work was supported by the DFG in the framework of the SFB910.

\end{document}